# PRELIMINARY MODELLING OF RADIATION LEVELS AT THE FERMILAB PIP-II LINAC

L. Lari, Fermilab, Batavia, IL, USA and CERN, Geneva, Switzerland
F. Cerutti, L. S. Esposito, CERN, Geneva, Switzerland
C. Baffes, S. J. Dixon, N. V. Mokhov, I. Rakhno, I. S. Tropin, Fermilab, Batavia, IL, USA

*Abstract*

PIP-II is the Fermilab's flagship project for providing powerful, high-intensity proton beams to the laboratory's experiments. The heart of PIP-II is an 800-MeV superconducting linac accelerator. It will be located in a new tunnel with new service buildings and connected to the present Booster through a new transfer line. To support the design of civil engineering and mechanical integration, this paper provides preliminary estimation of radiation level in the gallery at an operational beam loss limit of 0.1 W/m, by means of Monte Carlo calculations with FLUKA and MARS15 codes.

## INTRODUCTION

The Proton Improvement Plan II project, or PIP-II, encompasses a set of upgrades and improvements to the Fermilab accelerator complex aimed at supporting a world-leading neutrino program over the next several decades. As an immediate goal, PIP-II is focused on upgrades capable of providing 120-GeV proton beam power in excess of 1 MW on a target at the start of the Long-Baseline Neutrino Facility/Deep Underground Neutrinos Experiment (LBNF/DUNE) program, currently anticipated for the middle of 2020s. Moreover, PIP-II is a part of the long-term goal of establishing a high-intensity proton facility that is unique within the world, ultimately leading to multi-MW capabilities at Fermilab [1].

## PIP-II DESIGN UPGRADES

The central element of PIP-II is a new linac, constructed of CW-capable accelerating structures and CryoModules (CM). It accelerates H- up to 800 MeV in a sequence of normal conducting and superconducting accelerator sections (see Fig. 1). It is built to operate with an averaged H- beam current of 2 mA and a beam duty factor of 1.1%.

The PIP-II linac is followed by a stripping beam transfer line to bring the final proton beam to the Booster. Upgrades to a number of systems in the Booster, Recycler and Main Injector are required to accommodate the new higher injection energies and beam intensity in cascade in each machine, before protons reach the LBNF target facility.

## PIP-II CIVIL CONSTRUCTIONS

The new linac requires new constructions on Fermilab site to house the accelerator components and support equipment and to connect it to the existing Booster through the new beam transfer line (see Fig. 2).

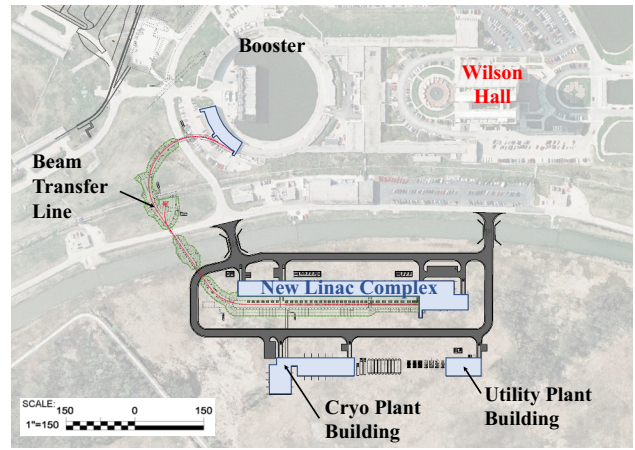

Figure 2: Plan view of PIP-II facilities. Location of the new services buildings, the new linac complex, including high bay building, tunnel & gallery, and bean transfer line are shown with respect to Fermilab Wilson Hall.

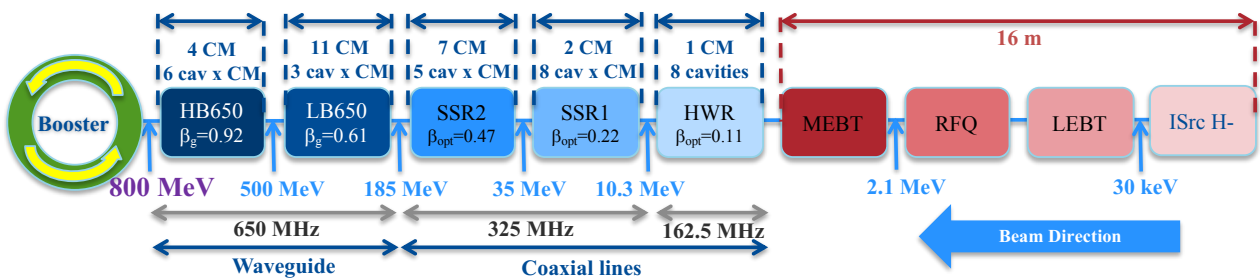

Figure 1: Schematic block diagram of the PIP-II accelerator lattice, from H- ions source (right) to the Booster connection (left). The blue items are superconducting (i.e. the Half Wave Resonator (HWR), the two Single Spokes Resonator sections (SSR1 & SSR2), the Low and High β elliptical CryoModules (LB650 and HB650)), while the front end is normal conducting. The optimal β are shown for HWR, SSR1 and SSR2 while the geometric β are shown for the LB650 and HB650. The operational frequencies for the superconducting section are also shown.

The new tunnel, a part of the new linac complex (see Fig. 2), is sized to accommodate the approximately 200 m superconducting section. It also includes space for a possible future upgrade up to 1.2 GeV, through the installation of four additional high β CryoModules. The warm front end is located in the high bay, a part of the new linac complex, at the entrance of the tunnel and a 210-m gallery will be constructed parallel to the below-grade linac tunnel to house the equipment needed to operate the beamline components. Penetrations between the tunnel and gallery will be provided for utilities, controls, cooling water, cryogens and related operational service.

## LINAC ENCLOSURE

Presently three different cross section designs are under study taking into account civil construction, mechanical integration and radiation constraints. The main penetrations between the PIP-II tunnel and gallery are supposed to house waveguide or coaxial lines for the powering of the Radio Frequency (RF) cavities. Additional penetrations are required for housing signal or general servicing cables as needed for the operation of the PIP-II accelerator (see Fig. 3).

## FLUKA CALCULATIONS

The FLUKA Monte Carlo code [2,3] is used for the evaluation of the prompt radiation in the gallery, assuming 0.1 W/m proton losses uniformly distributed along the superconducting section and taking into account the different beam energies along the machine. As a conservative approach the impact angle is set to 3 mrad with respect to the beam direction.

The level of 0.1 W/m uniform losses allowed for PIP-II accelerator, corresponds to set an activation level limit of 15 mrem/h at 30 cm from beam component surface. It has to be pointed out that this loss goal is set one factor lower with respect to other facilities, where this limit is 1 W/m for "hands-on" maintenance [4, 5].

For the evaluation of the effect of the different layouts on the dose in the gallery at the exit of penetrations, a simplified approach was used. It consists in distributing the beam losses on the internal surface of a standard stainless steel beam vacuum pipe of 2 mm thickness, at diameter sized for the low β and high β section.

## FLUKA RESULTS

Simulation results show that the best layout scenario for penetration is alternative A. Alternative B shows approximately twice the amount of dose, while C is about six time larger (see Fig. 4).

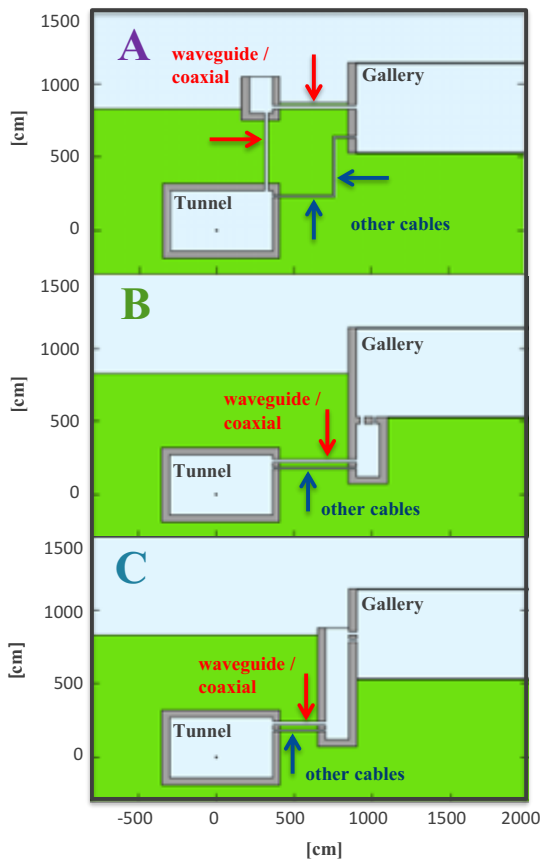

Figure 3: FLUKA models of alternatives A, B and C for penetrations. The model layout is built in FLUKA for radiation study purposes and based on civil engineering drawings. Location of waveguide/coaxial and other cables are shown for the different layouts. The average transverse aperture is 1000 $cm^2$ for waveguides, 500 $cm^2$ for coaxial lines and 100 $cm^2$ for other service cables.

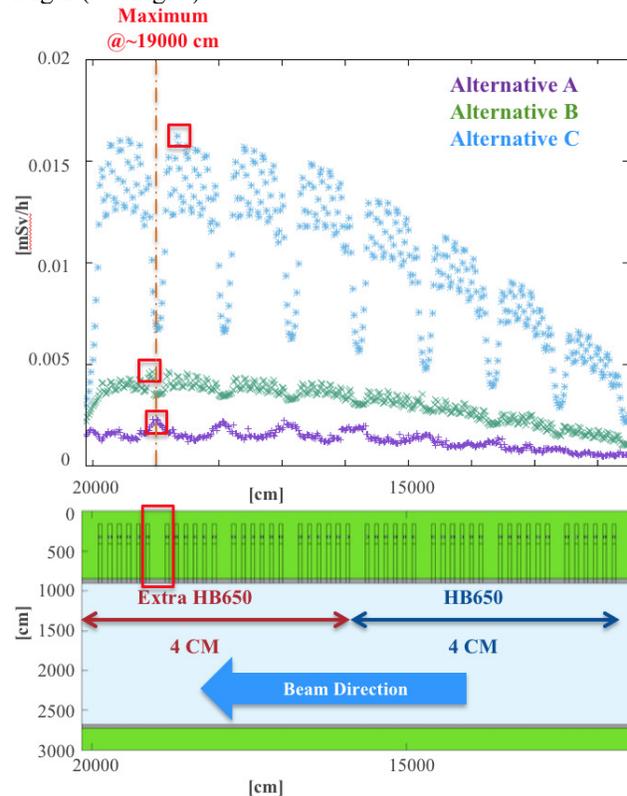

Figure 4: FLUKA averaged ambient dose equivalent rate results in the gallery, near the penetration exits fort the 3 alternatives (top plot). The maximum in the gallery is at the location in the tunnel situated at ~190 m from HWR in the extra high-β section (bottom plot).

For alternative B and C, the maximum is reached where a penetration is present, while for alternative A, the maximum dose is predicted in an area between the two-last extra high-β CryoModules where there are no penetrations (see Fig. 5). Simulations take into account data of Fermilab site soil composition and concrete used in recent laboratory facilities built at Fermilab.

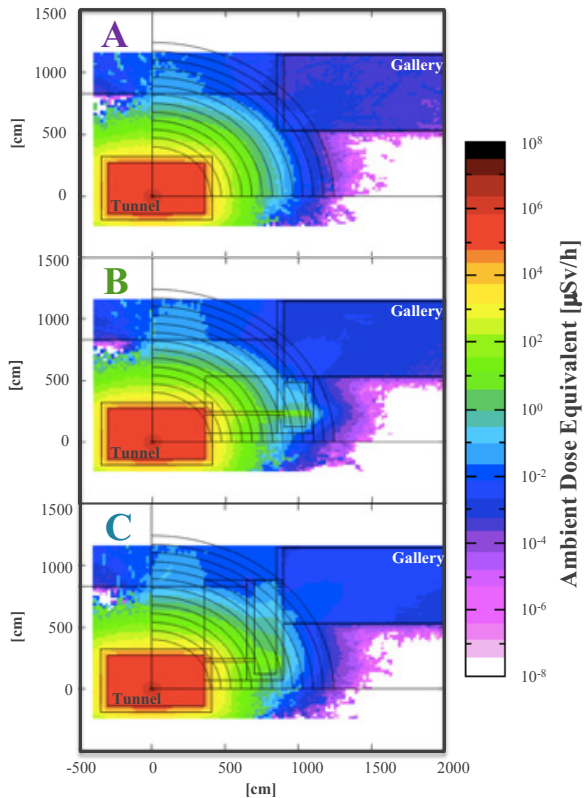

Figure 5: Ambient dose equivalent rates at the hot locations for the three cases: at ~190 m from HWR for alternative A, at ~188 m from HWR for B and at ~194 m from HWR for C (as shown in Fig. 4). It should be noted that for alternative A, the penetration layout geometry is not shown in the top plot, since the maximum is reached in the area where the penetration is not present. The quarter circles shown in the three plots represent fictitious surfaces used to implement a region importance biasing to compensate the attenuation of the neutron field from the tunnel to the gallery. Statistical uncertainties are approximately 15% for bin sizes of 400 cm$^2$ transverse and 40 cm along the tunnel length.

## EXTRAPOLATION BASED ON MARS15

The simplified approach used is valid for comparing the three alternatives in terms of dose released in the gallery, but it is not giving realistic results. Indeed, the presence of beam elements can bring different estimations. For this reason, a comparison with a similar linac, i.e. the European Spallation Source (ESS) one, was performed.

Effective dose MARS15 [6,7] simulation results applied to ESS linac [8] with a complete beam line geometry, accelerating fields in cavities and magnetic fields in quadrupoles give $0.8 \times 10^4$ mrem/h in the tunnel at the penetration entrance. Corresponding effective dose calculated in this study with FLUKA without the beamline elements is $2 \times 10^4$ mrem/h. Both values are given at the 800 MeV longitudinal position and are based on a proton loss rate of 0.1 W/m. All the complexity of the realistic MARS15 modelling gives a reduction in dose of 2.5 with respect to the simplified bare-pipe FLUKA results.

Based on the above MARS15 dose in the tunnel and using the analytical calculations applied to the alternative A layout, one finds that the dose at this location in the gallery is safely below the unlimited occupancy criterion of 0.25 mrem/h (see Fig.6) [9].

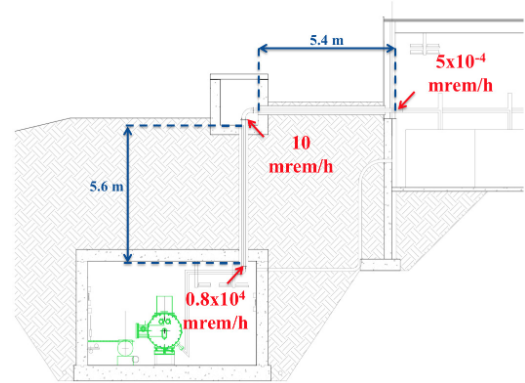

Figure 6: Alternative A cross section civil engineering drawing with values of effective dose shown at the exit/entrance of the tunnel, ground level (inside the waveguide vault) and gallery. The estimates are based on analytical calculations with universal curves of dose attenuation.

## CONCLUSIONS

The FLUKA results presented in this paper show that alternative A is the best choice to lower radiation in the PIP-II gallery. However civil engineering and mechanical constraints make the alternative B more suitable.

Some improvements such as extra shielding in the gallery penetration exit for the higher doses areas can bring the alternative B at the same level of radiation in the gallery as the alternative A.

Detailed studies will follow this preliminary evaluation by the implementation of a realistic model of the PIP-II machine in the simulations and taking into account the filling factors for the penetrations and the on-going optimization of the mechanical and civil constructions designs.

## ACKNOWLEDGMENTS

This document was prepared in part using the resources of the Fermi National Accelerator Laboratory (Fermilab), a U.S. Department of energy, Office of science, HEP User Facility. Fermilab is managed by Fermi Research Alliance, LLC (FRA), acting under Contract No. DE-AC0207CH11359.

This research also used a computing discretionary allocation at the Argonne Leadership Computing Facility, which is a DOE Office of Science User Facility supported under Contract DE-AC02-06CH11357.

The support of Roberto Cevenini with the CERN computing facility is warmly acknowledged.